# A Framework for the Interoperability of Cloud Platforms: Towards FAIR Data in SAFE Environments

Robert L. Grossman[1,2], Rebecca R. Boyles[3], Brandi N. Davis-Dusenbery[4], Amanda Haddock[5], Allison P. Heath[6], Brian D. O'Connor[7], Adam C. Resnick[6], Deanne M. Taylor[6,8], and Stan Ahalt[9],

January 22, 2024

1. Center for Translational Data Science, University of Chicago, Chicago IL USA
2. Corresponding author
3. RTI International, Research Triangle Park NC USA
4. Velsera, Charlestown MA USA
5. Dragon Master Initiative, Kechi KS USA
6. Children's Hospital of Philadelphia, Philadelphia PA USA
7. Nimbus Informatics, Carrboro NC USA
8. University of Pennsylvania Perelman School of Medicine, Philadelphia PA USA
9. University of North Carolina, Chapel Hill, Chapel Hill NC USA

# Abstract

As the number of cloud platforms supporting scientific research grows, there is an increasing need to support interoperability between two or more cloud platforms.  A well accepted core concept is to make data in cloud platforms Findable, Accessible, Interoperable and Reusable (FAIR).  We introduce a companion concept that applies to cloud-based computing environments that we call a **S**ecure and **A**uthorized **F**AIR **E**nvironment (SAFE).  SAFE environments require data and platform governance structures and are designed to support the interoperability of sensitive or controlled access data, such as biomedical data.   A SAFE environment is a cloud platform that has been approved through a defined data and platform governance process as authorized to hold data from another cloud platform and exposes appropriate APIs for the two platforms to interoperate.

# Background

As the number of cloud platforms supporting scientific research grows [1], there is an increasing need to support cross-platform interoperability.  By a cloud platform, we mean a software platform in a public or private cloud [2] for managing and analyzing data and other authorized functions.  With interoperability between cloud platforms, data does not have to be replicated in multiple cloud platforms but can be managed by one cloud platform and analyzed by researchers in another cloud platform.  A common use case is to use specialized tools in another cloud platform that are unavailable in the cloud platform hosting the data.  Interoperability also enables cross-platform functionality, allowing researchers analyzing data in one cloud platform to obtain the necessary amount of data required to power a statistical analysis, to validate an analysis using data from another cloud platform, or to bring together multiple data types for an integrated analysis when the data is distributed across two or more cloud platforms.  In this paper, we are especially concerned with frameworks that are designed to support the interoperability of sensitive or controlled access data, such as biomedical data or qualitative research data.

There have been several attempts to provide frameworks for the interoperating cloud platforms for biomedical data, including those by the GA4GH organization [3] and by the European Open Science Cloud (EOSC) Interoperability Task Force of the FAIR Working Group [4]. A key idea in these frameworks is to make data in cloud platforms findable, accessible, interoperable and reusable (FAIR) [5].

The authors have developed several cloud platforms operated by different organizations and were part of a working group, one of whose goals was to increase the interoperability between these cloud platforms. The challenge is that even when a dataset is FAIR and in a cloud platform (referred to here as Cloud Platform A), in general the governance structure put in place by the organization sponsoring Cloud Platform A (called the Project Sponsor below) requires that sensitive data remain in the platform and only be accessed by users within the platform. Therefore, even if a user was authorized to analyze the data, there was no simple way for the user to analyze the data in any cloud platform (referred to here as Cloud Platform B), except for the single cloud platform operated by the organization (Cloud Platform A).

There are several reasons for this lack of interoperability between cloud platforms hosting sensitive data: First, as just mentioned, for many cloud platforms, it is against policy to remove data from the cloud platform; instead, data must be analyzed within the cloud platform.

Second, in some cases, to manage the security and compliance of the data, often there is only a single cloud platform that has the right to distribute controlled access data; other cloud platforms may contain a copy of the data, but by policy cannot distribute it.

Third, a typical clause in a data access agreement requires that if the user elects not to use Cloud Platform A, the user's organization is responsible for assessing and attesting to the security and compliance of Cloud Platform B. This can be difficult and time consuming unless there is a pre-existing relationship.

Fourth, once a Sponsor has approved a *single* cloud platform as authorized to host data and to analyze the hosted data, there may be a perception of increased risk to the Sponsor in allowing other third party platforms to be used to host or to analyze the data. Because of this increased risk, there has been limited interoperability of cloud platforms for controlled access data.

The consensus from the working group was that interoperability of data and an acceleration of research outcomes could be achieved if standard interoperating principals and interfaces could describe which platforms had the right to distribute a dataset and which cloud platforms could be used to analyze data.

In this note, we introduce a companion concept to FAIR that applies to cloud-based computing environments that we call a **S**ecure and **A**uthorized **F**AIR **E**nvironment (SAFE). The goal of the SAFE framework is to address the four issues described above that today limit the interoperability between cloud platforms. The cloud-based framework consisting of FAIR data in SAFE environments is intended to apply to research data that has restrictions on its access or its distribution or both its access and distribution. Some examples are: biomedical data [3], [6], including EHR data, clinical/phenotype data, genomics data, imaging data; social science data [7] and administrative data [8]. We emphasize that the environment itself is not FAIR in the sense of [5], but rather that a SAFE environment contains FAIR data and is designed to be part of a framework to support the interoperability of FAIR data between two or more data platforms.

Also, SAFE cloud platforms are designed to support *platform* governance decisions about whether data in one cloud platform may be linked or *transferred* to another cloud platform, either for direct use by researchers or to redistribution. As we will argue below, SAFE is designed to support decisions between two or more cloud platforms to interoperate in the sense that data may be moved between them, but is not designed nor intended to be a security or compliance level describing a single cloud platform.

The proposed SAFE framework provides a way for a Sponsor to "extend its boundary" to selected third party platforms that can be used to analyze the data by authorized users.  In this way, researchers can use the platform and tools that they are most comfortable with.

In order to discuss the complexities of an interoperability framework across cloud based resources, in the next section, we first define some important concepts from data and platform governance.

# Distinguishing Data and Platform Governance

We assume that data is generated by research projects and that there is an organization that is responsible for the project.  We call this organization the *Project Sponsor.* This can be any type of organization, including a government agency, an academic research center, a not-for-profit organization, or a commercial organization.

In the framework that we are proposing here, the Project Sponsor sets up and operates frameworks for 1) data governance and 2) platform governance.  The Project Sponsor is ultimately responsible for the security and compliance of the data and of the cloud platform. *Data governance* includes: approving datasets to be distributed by cloud platforms, authorizing users to access data, and related activities.  *Platform governance* includes: approving cloud platforms as having the right to distribute datasets to other platforms and to users and approving cloud platforms as authorized environments so that the cloud platforms can be used by users to access, analyze, explore datasets.

By *controlled access data*, we mean data that is considered sensitive enough that agreements for the acceptable use of the data must be signed.  One between the organization providing the data (the *Data Contributor*) and the Project Sponsor and another between researchers (which we call *Users* in the framework) accessing the data and the Project Sponsor.  Controlled access data arises, for example, when research participants contribute data for research purposes through a consent process, and a researcher signs an agreement to follow all the terms and conditions required by the consent agreements of the research participants or by an Institutional Review Board (IRB) that approves an exemption so that consents are not required.

Commonly used terms that are needed to describe SAFE are contained in Table 1. Table 2 describes the roles and responsibilities of the Project Sponsor, Platform Operator, and User.

As is usual, we use the term *authorized user,* as someone who has applied for and been approved for access to controlled-access data. See Table 1 for a summary of definitions used in this paper.

One of the distinguishing features of our interoperability framework is that we formalize the concept of an authorized environment. An *authorized environment* is a cloud platform workspace or computing / analysis environment that is approved for the use or analysis of controlled access data.

Using the concepts of authorized user and authorized environment, we provide a framework enabling the interoperability between two or more cloud platforms.

| Term | Definition |
|---|---|
| authorized environment | An authorized environment is a cloud platform workspace or computing infrastructure that is approved for the use or analysis of controlled access data. |
| authorized platform identifier (APID) | SAFE assumes that cloud platforms have a globally unique identifier (GUID) identifying them called the authorized platform identifier (APID). |
| authorized platform network (APN) and authorized platform network identifier (APNI) | SAFE assumes that cloud platforms form networks consisting of 2 or more cloud platforms called authorized platform networks (APN) that are identified by a globally unique identifier called an authorized platform network identifier (APNI). |
| authorized region ID | SAFE assumes that geographic regions are identified by a globally unique identifier called an Authorized Region ID (ARID). |
| authorized user | An authorized user is someone who has applied and been approved for access to controlled-access data. |
| controlled access data | Controlled access data is sensitive enough that agreements for the acceptable use of the data must be signed between the organization providing the data (the Data Contributor) and the Project Sponsor and between researchers (which we call |

|  | Users in the framework) accessing the data and the Project Sponsor. |
|---|---|
| cloud platform | A software platform in a public or private cloud for managing, analyzing and sharing data, as well as other authorized functions. |
| data access agreement and data contributor agreement | We use the term Data Access Agreement for the agreement signed by the researcher, or, in some cases, the researcher's organization, and the term Data Contributor Agreement for the agreement signed by the individual or organization that contributes the data to be distributed to the community. In the Data Access Agreement, the user, or user's organization, commonly agrees to protect the privacy and confidentiality of the data and adhere to any terms and conditions that are specified in the Data Contributor's Agreement [6]. In particular, the Data Access Agreement requires the user to analyze the data in a secure environment in order to protect the data and agree not to attempt to re-identify any deidentified data. The Data Contributor's Agreement may include what are sometimes called secondary use restrictions, such as restrictions precluding commercial use of the data [7], and such restrictions flow through to the Data Access Agreement. |
| open access data | By open access data, we mean data that is not sensitive enough to require the special protections of controlled access data. Note that open access data might impose some restrictions, such as requiring the user to acknowledge the source of the data in any publications or talks. |
| data governance | Data governance includes: approving datasets to be distributed by cloud platforms, authorizing users to access data, and related activities. |
| platform governance | Platform governance includes approving cloud platforms as having the right to distribute datasets to other platforms and to authorized users on the platforms and approving cloud platforms as authorized environments so that the cloud platforms can be used by authorized users to access, analyze, and explore datasets |
| right to distribute | The Project Sponsor associated with a dataset may authorize a cloud platform as having the right to distribute selected datasets to authorized users in other authorized environments that have been approved by the Project Sponsor. |

Table 1. Definitions of key terms related to data and platform governance supporting SAFE.

# SAFE Environments

Below we describe some suggested processes for authorizing environments, including having their security and compliance reviewed by the appropriate official or committee determined by the platform governance process.   We also argue that the environments should have APIs so that they are findable, accessible and interoperable, enabling other cloud platforms to interoperate with it.  As mentioned above, we use the acronym SAFE for **S**ecure and **A**uthorized **F**AIR **E**nvironments to describe these types of environments. In other words, a SAFE environment is a cloud platform that has been approved through a platform governance process as an authorized environment and exposes an API enabling other cloud platforms to interact with it (Figure 1).

In this paper, we make the case that SAFE environments are a natural complement to FAIR data **and establishing a trust relationship between a cloud platform with FAIR data and a cloud platform that is a SAFE environment for analyzing data is a good basis for interoperability.** Examples of the functionality to be exposed by the API and proposed identifiers are discussed below. Importantly, our focus is to provide a framework for attestation and approvals to support interoperability. Definition of the exact requirements for approvals is based on the needs of a particular project sponsor and out of scope of this manuscript.

Of course, a cloud platform can include both FAIR data and a SAFE environment for analyzing data.  The issue of interoperability between cloud platforms arises when a researcher using a cloud platform that is a SAFE environment for analyzing data needs to access data from another cloud platform that contains data of interest.

We emphasize that the framework applies to all types of controlled-access data, (e.g., clinical, genomic, imaging, environmental, etc.) and that decisions about authorized users and authorized platforms depend upon the sensitivity of the data, with more conditions for data access and uses as the sensitivity of the data increases.

The SAFE framework that we are implementing uses the following identifiers:

1. SAFE assumes that cloud platforms have a globally unique identifier (GUID) identifying them, which we call an *authorized platform identifier (APID)*.
2. SAFE assumes that cloud platforms form networks consisting of 2 or more cloud platforms, which we call *authorized platform network (APN).*  Authorized platform networks have a globally unique identifier, which we call an *authorized platform network identifier (APNI)*. As an example, cloud platforms in an authorized platform network can sign a common set of agreements or otherwise agree to interoperate. A particular cloud platform can interoperate with all or selected cloud platforms in an authorized platform network.
3. SAFE assumes that geographic regions are identified by a globally unique identifier, which we call an Authorized Region ID (ARID). For example, the entire world may be an authorized region, or a single country may be the only authorized region. SAFE assumes that datasets that limit their distribution and analysis to specified regions identify these regions in their metadata.

To implement SAFE, we propose that a cloud environment support an API that exposes metadata with the following information:

- Authorized Platform Identifier (APID)
- A list of the Authorized Platform Network Identifiers (APNIs) that it belongs to.

A particular authorized platform network must also agree to a protocol for securely exchanging the APID and list of APNIs that it belongs to, such as transport layer security (TLS) protocol.

In addition, cloud platforms that host data that can be accessed and analyzed in other cloud platforms, should associated with each dataset metadata that specifies: a) whether the data can be removed from the platform (i.e. does the platform have the right to distribute data); b) a list of authorize platform networks that have been approved as authorized environments to access and analyze the data; and, c) an optional list of authorized region IDs (ARIDs) describing any regional restrictions on where the data may be accessed and analyzed.

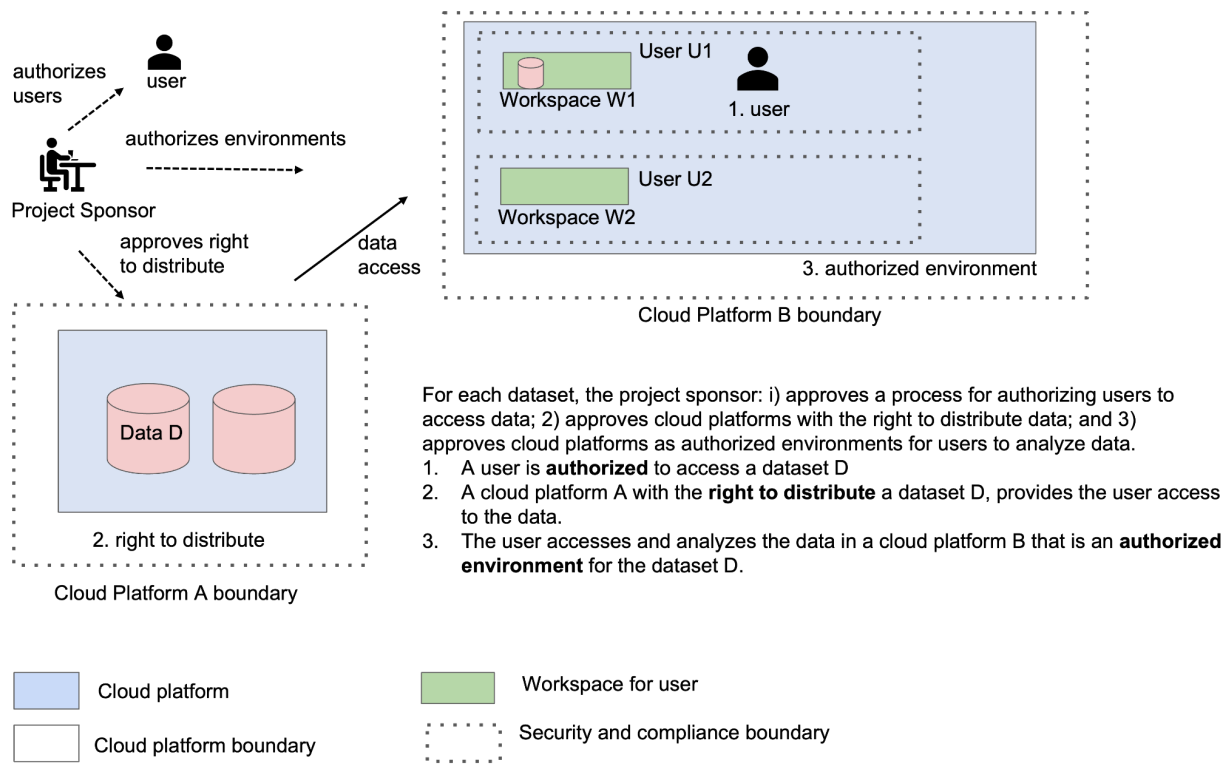

Figure 1. An overview of supporting FAIR data in SAFE environments.

# Platform Governance

## Examples of Platform Governance Frameworks

An example of a process for authorization in an environment is provided by the process used by the NIH Data Access Committees (DACs) through the dbGaP system [9] for sharing genomics data [10].  Currently, if a NIH DAC approves a user's access to data, and if the user specifies in the data access request (DAR) application that a cloud platform will be used for analysis, then the user's designated IT Director takes the responsibility for a cloud platform as an authorized environment for the user's analysis of controlled access data, and a designated official at the user's institution (the Signing Official) takes the overarching responsibility on behalf of the researcher's institution.

As another example, the platform governance process may follow the "NIST 800-53 Security and Privacy Controls for Information Systems and Organizations" framework developed by the US National Institute for Standards and Technology (NIST) [11]. This framework has policies, procedures, and controls at three Levels - Low, Moderate and High, and each organization designates a person that can approve an environment by issuing what is called an Authority to Operate (ATO). More specifically, in this example, the platform governance process may require the following to approve a cloud platform as an authorized environment for hosting controlled access data: 1) a potential cloud platform implement the policies, procedures and controls specified by NIST SP 800-53 at the Moderate level; 2) a potential cloud platform have an independent assessment by a third party to ensure that the policies, controls and procedures are appropriately implemented and documented; 3) an appropriate official or committee evaluate the assessment, and if acceptable, approves the environment as an authorized environment by issuing an Authority to Operate (ATO) or following another agreed to process;   4) yearly penetration tests by an independent third party, which are reviewed by the appropriate committee or official.

Many US government agencies follow NIST SP 800-53, and a designated government official issues an Authority to Operate (ATO) when appropriate after the evaluation of a system [11]. In the example above, we are using the term "authority to operate" to refer to a *more* general process in which *any organization* decides to evaluate a cloud platform using *any* security and compliance framework and has completed all the steps necessary so that the cloud platform can be used.  In the example, an organization, which may or may not be a government organization, uses the NIST SP 800-53 security and compliance framework and designates an individual within the organization with the role and responsibility to check that 1), 2) and 4) have been accomplished and issues an ATO when this is the case.

## The Right to Distribute Controlled Access Data

 In general, when a user or a cloud platform is granted access to controlled access data, the user or platform does not have the right to redistribute the data to other users, even if the other user has signed the appropriate Data Access Agreements. Instead, to ensure there is the necessary security and compliance in place, any user accessing data as an authorized user must access the data from a platform approved for this purpose.  We refer to platforms with the ability to share controlled access data in this way as having the *right to distribute* the authorized data.

One of the core ideas of SAFE is that data which has been approved for hosting in a cloud platform can be accessed and transferred to another cloud platform in the case that: the first cloud platform has the right to distribute the data and the second cloud platform is recognized as an authorized environment for the data following an approved process, such as described in the next section.  There remains the possibility that the cloud platform requesting access to the data is in fact an imposter and not the authorized environment it appears to be.  For this reason, as part of SAFE, we recommend that the cloud platform with the right to distribute data should verify through a chain of trust that it is indeed the intended authorized environment.

## Basis for Approving Authorized Environments

The guiding principle of SAFE is that research outcomes are accelerated by supporting interoperability of data across authorized environments. While the specific requirements may vary by project and project sponsor, in order to align with this principle, it is critical that Project Sponsors define requirements transparently and support interoperability when the requirements are met.

Below we provide examples of approaches and requirements project sponsors may use in approving an Authorized Environment. As mentioned above, NIST SP 800-53 provides a basis for authorizing an environment, but there are many frameworks for evaluating the security and compliance of a system that may be used. As an example, the organization evaluating the cloud platform may choose to use a framework such as NIST SP 800-171 [12], or may choose another process for approving a cloud platform as an authorized environment rather than issuing an ATO.

For example, both the Genomic Data Commons [6] and the AnVIL system [13] follow NIST SP 800-53 at the Moderate Level and the four steps described above. The authorizing official for the Genomic Data Commons is a government official at the US National Cancer Institute, while the authorizing official for AnVIL is an organizational official associated with the Platform Operator.

Two or more cloud platforms can interoperate when both the Sponsors and Operators each agree to: 1) use the same framework and process for evaluating cloud platforms as authorized environments; 2) each authorize one or more cloud platforms as authorized environments for particular datasets; 3) each agree to a common protocol or process for determining when a given cloud platform is following 1) and 2). Sometimes, this situation is described as the platforms having a *trust relationship* between them.

## Basis for Approving the Right to Distribute Datasets

For each dataset, a data governance responsibility is to determine the right of a cloud based data repository to distribute data to an authorized user in an authorized environment. To reduce risk of privacy and confidentiality breach, the data governance process may choose to limit the number of data repositories that can distribute a particular controlled access dataset and to impose additional security and compliance requirements on those cloud based data repositories that have the right to distribute particular sensitive controlled-access datasets. These risks of course must be balanced with the imperative to accelerate research and improve patient outcomes which underlies the motivations of many study participants.

### Interoperability

SAFE is focused on the specific aspect of interoperability of whether data hosted in one cloud platform can be analyzed in another cloud platform.

With the concepts of an authorized user, an authorized environment, and the right to distribute, interoperability is achieved when two or more cloud platforms have the right to distribute data to an authorized user in a cloud based authorized environment.

This suggests a general principle for interoperability: **the data governance process for a dataset should authorize users, the platform governance process for a dataset should authorize cloud platform environments, and two or more cloud platforms can interoperate by trusting these authorizations.**

Figure 2 summarizes some of the key decisions enabling two cloud platforms to interoperate using the SAFE framework.

## Towards FAIR Data in SAFE Environments

Today there are a growing number of cloud platforms that hold biomedical data of interest to the research community, a growing number of cloud-based analysis tools for analyzing biomedical data, and a growing challenge for researchers to access the data they need, since often the analysis of data takes place in a different cloud platform than the cloud platform that hosts the data of interest.

We have presented the concept of cloud-based authorized environments that are called SAFE environments, which are secure and authorized environments that are appropriate for the analysis of sensitive biomedical data.  The role of platform governance is to identify the properties required for a cloud platform to be an authorized environment for a particular dataset and to approve a cloud based platform that holds controlled access data to distribute the data to specific authorized platforms.

By standardizing the properties to be a SAFE environment and agreeing to the principle that the data governance process for a dataset should authorize users
and
the platform governance process should authorize cloud platform environments, then all that is required for two or more cloud platforms to interoperate is for the cloud platforms to trust these authorizations. We can shorten this principle to: "authorize the users, authorize the cloud platforms, and trust the authorizations." This is the core basis for interoperability in the SAFE framework. See Table 3 for a summary.

This principle came out of the NIH NCPI Community and Governance Working Group and is the basis for the interoperability of the data platforms in this group. We are currently implementing APID, APNI and AIRD identifiers as described above, as well as the dataset metadata describing whether a dataset can be redistributed or transferred to other data platforms for analysis.

| Name of Role | Scope of Role | Relevant Agreements |
| --- | --- | --- |
| Project Sponsor | The organization that sets up the governance frameworks and is ultimately responsible for data governance and platform governance, including all necessary security and compliance. | The Project Sponsor sets up frameworks and associated agreements to: 1) authorize users to access data. 2) authorizes cloud platforms to distribute data; and, 3) approves cloud platforms as authorized environments for users to access and analyze data. |
| Data Contributor | An organization that is providing one or more datasets to the Project Sponsor with the intention that it will be provided to the Platform Operator so that the dataset is available to the research community. | The Data Contributor Agreement is an agreement between the Data Contributor and the Project Sponsor governing the terms and conditions of the data, including any restrictions on its use. |
| Platform Operator | A platform operator is responsible for operating cloud platforms that provide access to datasets and to analysis tools and implementing controls that balance the availability of data with the obligation to protect the confidentiality and security of the data. | 1) There is an agreement between the Project Sponsor and a Platform Operator that sets the terms and conditions that enable the cloud platform to distribute data to authorized users and to transfer the data to authorized platforms (Right to Distribute Data Agreement). 2) There is an agreement between the Project Sponsor and a Platform Operator that sets the terms and conditions that enable the cloud platform to provide an environment to users so they can access and |

| | | analyze data (Authorized Environment Agreement). |
|---|---|---|
| User | A researcher using a cloud platform to access, explore, or analyze data. | 1) There is an agreement between a Project Sponsor and User setting the terms and conditions for the User to access and analyze data (Data Use Agreement). 2) There is sometimes an agreement between the User and Platform Operator governing the terms and conditions of the researcher's use of the cloud platform. |

Table 2. Key roles and responsibilities in the SAFE Framework.

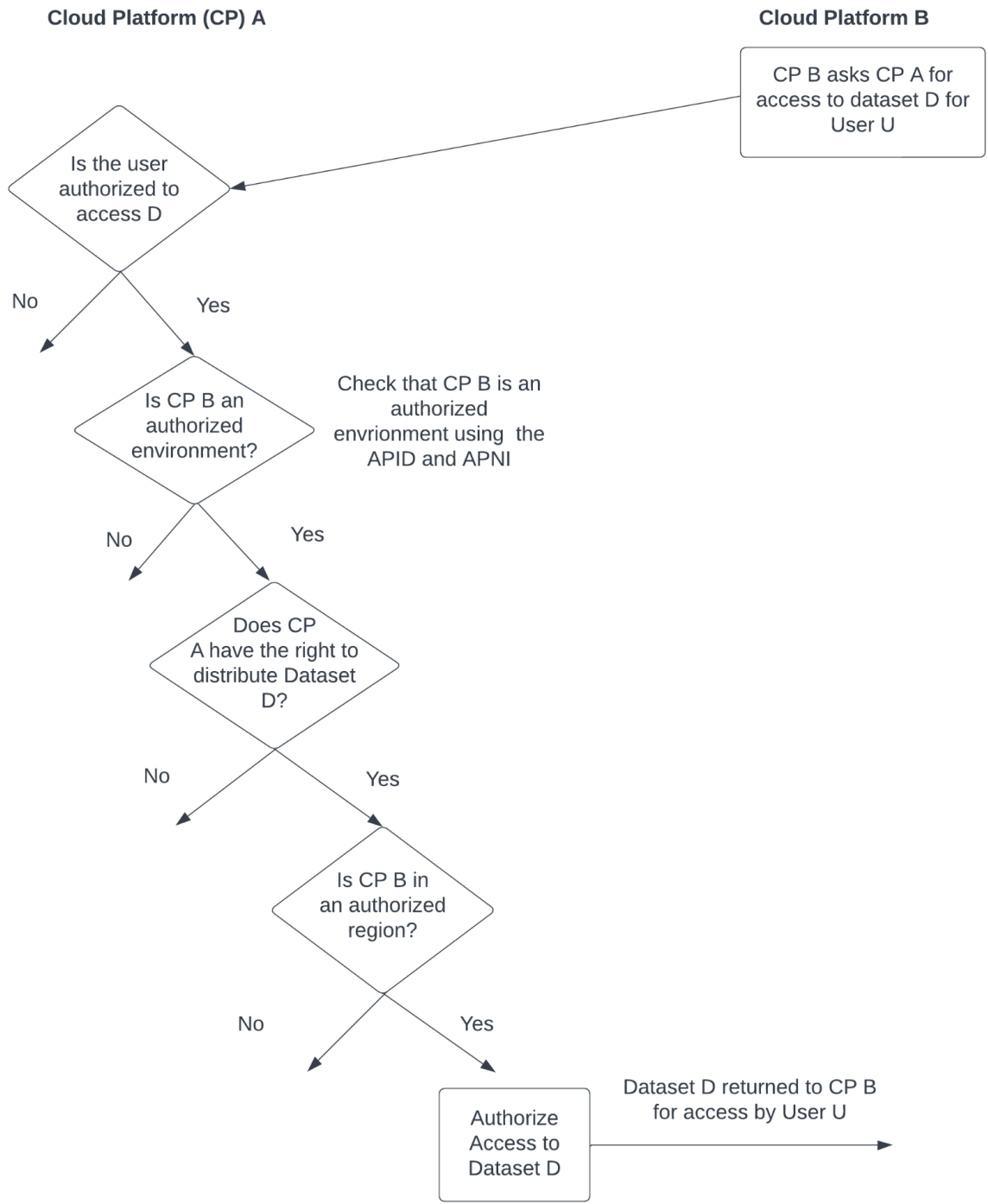

Figure 2. Some of the key decisions for interoperating two cloud platforms using the SAFE framework.

|  | **FAIR** | **SAFE** |
|---|---|---|
| Key concern | Datasets | Data platforms |
| Who is responsible | Data curators and research leads | Data platform operators and those that authorize data platforms to operate |
| Consensus process | Community consensus around data (data elements, controlled vocabularies, ontologies, etc. to support making data interoperable and reusable) and software services and APIs for finding and accessing data | Community consensus around when two data platforms can interoperable, including when one data platform (Platform A) can distribute a FAIR dataset to another data platform (Platform B) |
| Supporting services and APIs | Data services and APIs for finding and accessing datasets | Data services and APIs so that one data platform can trust another data platform and exchange datasets, containers, and other objects with it. |

Table 3. A summary of some of the important distinctions between FAIR data and SAFE platforms. FAIR is concerned with datasets and how they can be discovered, accessed, and semantically and syntactically interoperate. SAFE is concerned with data platforms containing FAIR datasets and when the decisions of their owners/operators and the governance, privacy and security policies they individually follow allow them to interoperate in the narrow sense of when data in one of the platforms can be transferred to the other for and accessed by an authorized user.

**Acknowledgements and Disclaimer**

This document captures discussions of the NIH Cloud-Based Platform Interoperability (NCPI) Community/Governance Working Group that have occurred over the past 24 months, and we want to acknowledge the contributions of this working group. This working group included personnel from federal agencies, health systems, industry, universities, and patient advocacy groups. However, this document does not represent any official decisions or endorsement of potential policy changes and is not an official work product of the NCPI Working Group. Rather, it is a summary of some of the working group discussions and is an opinion of the authors.

**Author Contributions**

All the authors contributed to the drafting and review of the manuscript.

**Competing interests**

One of the authors (BND-D) is an employee of a for-profit company (Velsera).

**Funding**

Research reported in this publication was supported in part by the following grants and contracts: the NIH Common Fund under Award Number U2CHL138346, which is administered by the National Heart, Lung, and Blood Institute of the National Institutes of Health; the National Heart, Lung, and Blood Institute, National Institutes of Health, Department of Health and Human Services under the Agreement No. OT3 HL142478-01 and OT3 HL147154-01S1; National Cancer Institute, National Institutes of Health, Department of Health and Human Services under Contract No. HHSN261201400008C; and ID/IQ Agreement No. 17X146 under Contract No. HHSN261201500003I. The content is solely the responsibility of the authors and does not necessarily represent the official views of the National Institutes of Health.